\documentclass[aps,prb,showpacs,superscriptaddress,groupedaddress]{revtex4-1}
\usepackage{graphicx}  
\usepackage{amsmath}
\usepackage{dcolumn}   
\usepackage{bm}        
\usepackage{amssymb}   
\usepackage{epsfig}

\begin{document}
\title{The possibility to predict crack patterns on dynamic fracture}
\author{Lucas M\'{a}ximo Alves}
\email{lucasmaximoalves@gmail.com}
\affiliation{GTEME - Grupo de Termodin\^{a}mica, Mec\^{a}nica e Eletr\^{o}nica dos Materiais, Departamento de Engenharia de Materiais, Setor de Ci\^{e}ncias Agr\'{a}rias e de Tecnologia, Universidade Estadual de Ponta Grossa, Av. Gal. Carlos Calvalcanti, 4748, Campus UEPG/Bloco CIPP, Uvaranas, Ponta Grossa-PR, Brasil, CEP. 84030.900, 84031-510 }
\author{Rui F. R. M. Lobo}
\email{rfl@fct.unl.pt}
\affiliation{Nanophysics and Energy Efficiency Group (GNCN), Center of Technology and Systems (CTS), Departamento de Fisica, Faculdade de Ci\^{e}ncias e Tecnologia, Universidade Nova de Lisboa, 2829-516 Caparica, Portugal }
\date{\today}

\begin{abstract}
The Maximum Energy Dissipation Principle (MEDP) for dynamics fracture, far from equilibrium, proposed by Slepyan was modified. This modification includes a decoupling between the injected and dissipated energy by adding of a time delay and a description of the ruggedness produced by dissipation patterns. A time delayed energy conservation equation is deduced and dynamical equations that describe the dynamical system evolution were obtained in analogous way to the Slepyan's calculations. The conditions for the rising of the instability process were presented by a bifurcation map. These results shown that the theoretical framework proposed can describe the instability process along with dissipation patterns formation. This proposal was applied to dynamics fracture where it was possible to explain the results obtained by Fineberg-Gross for the fast crack propagation in PMMA. For unstable or dynamical crack propagation it is shown the possibility to predict crack patterns using this MEDP for other experimental configurations.
\end{abstract}
\maketitle

\section{\label{sec:1}Introduction}
Some mechanical systems display instability behavior along with the phenomena of dissipative structures and patterns formation\cite{1,2,3,4}. It has been found that this occurs as a response to extreme energy input events, and cracks formation is a good illustration for that purpose\cite{4,5}. These structures and patterns typically exhibit a branched geometry that can be characterized as a fractal, between a finite range of scales $\epsilon_{min}\leq \epsilon \leq \epsilon_{max})$\cite{3,4}.  However, there is not yet a complete treatment in Mechanics and Non-Equilibrium Thermodynamics able to describe all this phenomenology, from the unstable dynamic until to the growth of a dissipation structure or the fractal pattern formation\cite{5,6,7,8,9}. From the basic assumption that there is a delay of rates and speeds between the system input and output, a principle of maximum energy dissipation was postulated and corresponding dynamic evolution equations were obtained. The dynamics of the system is described by these equations, from the generated instability to the formation of structures and fractal patterns. It was observed that the delay of those speeds gives rise to the appearance of iterated equations which in turn lead to the logistic maps which describe the evolution dynamics of the system. An application to dynamic fracture was implemented and the obtained analytical results were compared with the already observed experimental instabilities in the phenomenon of fast cracking. Furthermore it is concluded that the above proposed delay of velocities in the description of the instability is adequate to describe the process of energy dissipation, leading naturally to the mathematical description of fractal patterns formation, and explaining the fracture instability in high speed regime. This makes an essential and unifying step in physics of fracture towards solving a critical outstanding problem.

\section{\label{sec:2}Theoretical Framework}
It is known the existence of dissipation patterns in Nature and particularly well on fracture of materials \cite{1,2,3}. Over the years of research on Physics of Fracture it was realized that in the processes of unstable crack propagation where instability and catastrophic situations occur, structures and branched patterns are produced, which can only be described by a suitable dynamic theory involving the formation of fractals or multifractals. But a fundamental question is the following: Is it possible to predict the crack patterns on fast crack growth? In a branched crack, the number of elemental structure, $N_{s}\approx \left(\frac{l_{\min}}{L_{s}}\right)^{-D_{q}}$ , with a unitary energy, $u_{s}$ given by the spatial density, $\rho_{s}\equiv dN_{s}/dL_{s}$  , arises as a consequence of an efficient energy dissipation process searching alternatives modes of dissipation, where this number of minimal cracks is optimized by maximizing the dissipation and minimizing the energy transport \cite{4}. To answer that former question, one needs to consider the possibility to include together in the same theoretical framework: (i) the non-linear equations of a dynamical system; (ii) a fundamental thermodynamical dissipation principle to describe the instability process able to generate equations that supply iterated maps which can finally represent a pattern. Thus with this purpose one can use the dynamic fracture and add a modified version of the maximum dissipation principle (MEDP) proposed by Slepyan to forecast the experimental results found by Fineberg and co-workers \cite{1,2}.

The motivation of this work is based on the present absence of a non-equilibrium description for the complete phenomenological sequence, from the dynamics of instability to the growth of the dissipation structure or formation of fractal patterns. In order to contribute for filling such gap, and leading to a clarification of the involved thermodynamics, equations on maximum dissipation energy in fracture dynamics are developed.

\subsection{\label{subsec:1}The proposal of a modified Maximum Energy Dissipation Principle}
One assumes that the existence of a dissipation process with a branched pattern must be due to some non-equilibrium thermodynamic principle valid when instability conditions and the formed structure can be related. The first condition necessary is the presence of an excess of energy injected with regard to additional internal energy that the system can support. Another condition is the extremization of the energy distributed by the structure of the dissipation pattern.

A modified version of the Slepyany's maximum dissipation principle \cite{5,6} can be proposed from the balance of excess energy,
$\varphi$ between the supplied energy $\dot\phi =\vec G (L_{\phi}(t),v_{\phi}(t)).\vec v_{\phi}(t)$ and the dissipated energy as a crack growth, $\psi_{s}=\vec R_{s}(t').\vec v_{s}(t')$ where $\vec R_{s}(t')=u_{s}\vec \rho_{s}(t')$ is the resistance to growth the crack pattern, as follow:
\begin{eqnarray}
&&\delta\left[\varphi(L,\vec v)\right ]=\delta\int M(L,\vec v)ds \nonumber \\
&& =\int\delta\left [\vec G (L_{\phi}(t),\vec v_{\phi}(t)).\vec v_{\phi}(t) - \vec R_{s}(t').\vec v_{s}(t')\right]ds = 0 \nonumber \\ \label{(1)}
\end{eqnarray}
where $\Gamma$ is the curve that involves the structure generated in different steps of iteration, resulting
possibly in invariant structure by scale transformation. $G (L_{\phi}(t),\vec v_{\phi}(t))$ is the elastodynamic energy released rate and $L_{\phi}(t),\vec v_{\phi}(t)$  are the length and the velocities of the deformations imposed to the dynamical system, and $L_{\phi}(t'),\vec v_{\phi}(t')$ are the length and the velocities to form the dissipation pattern. One can admit that the decoupling between these two velocities as responsible by instability process, is given by a time delay $\tau$  between them. This time delay can be calculated by the input and the output energy of the system, in terms of ratio between the minimal characteristic crack length, $l_{min}$  and the maximal possible crack growth velocity, $\vec v_{\phi_{\max}}$ as:
\begin{equation}
\tau = \frac{l_{min}}{v_{\phi_{\max}}} = \frac{G.l_{min}}{G.c_{R}-R.v_{s}}=\frac{l_{min}}{[c_{R}-v_{s}/\alpha]}=characteristic \label{(2)}
\end{equation}
where $v_{\phi_{max}}\approx (2/3)c_{R}$ and $c_{R}$ is the limiting velocity given by Rayleigh waves velocities.  In addition one can define a parameter $\alpha = G/R$ which is  taken from Griffith-Irwin theory of fracture mechanics. Where $G$  is the elastic energy released rate and $R$  is the crack growth resistance.

Considering the existence of this time delay corroborated by the experimental correlations measurements between the oscillations in the crack growth velocity and the surface profile, of approximately $3.0\mu s$ and $1.0\mu s$  for PMMA and for the glass soda-lime, respectively \cite{1,2,3}, one can obtain from the variational calculation of the Eq. (\ref{(1)}), the following form for energy conservation principle:
\begin{equation}
\dot \phi \left( L_{\phi }(t),\vec v_{\phi}(t)\right) = \psi _{s}\left( L_{s}(t'),\vec v_{s}(t')\right) \label{(3)}
\end{equation}
This equation relates the injected power with the dissipated power.

In the experiments performed by Fineberg-Gross \cite{1,2} they showed that as the crack speed reaches a critical value a strong temporal correlation between velocity, $v_{\phi}(t)$, and the response in the form of fracture surface at $v_{\phi}(t')=dA_{\phi}(t + \tau)/dt$, takes place (having its notation changed, in the present text, to $L_{\phi}(t)$ instead $A_{\phi}(t)$ to designate the fracture surface length). The time delay measured between these two magnitudes present a value about of $3.0 \mu s$ for PMMA and $1.0 \mu s$ for soda-lime glass, for example, showing that there is a characteristic value for each material. By other side, Sharon et. al \cite{3} states that, after a time of $1.0\mu s$, the stress field is recuperated equal to the field stress of a single crack: "Thus $1.0\mu s$ after the death of a side branch of length $1.0mm$, the stress field throughout the singular zone will be that a single crack". Also Washabaugh and Knauss \cite{7} affirms that: "The generation of micro cracks, and thus, presumably, the retardation of crack propagation may be demonstrated by generating a material with controllable cohesive strength (a weak plane) while maintaining all other physical properties".

\subsection{\label{subsec:2}The Instability Dynamics Equations}
Therefore one assumes that in fracture dynamics the instability appears as a consequence of a time delay between the injected energy flux $\dot \phi(t)$  and the dissipated energy flux $\psi_{s}(t+\tau)$, in the form of fracture surfaces, as the injected energy flux interacts with the dissipated energy flux $\psi_{s}(t+\tau)$ after a time delay, given by the relaxation process in the material.

The remaining mathematical consideration of the variational principle from the Eq. (\ref{(1)})  supplies the following differential equations:

i) for the conditions $\vec v_{s}(t) = \vec v_{\phi }(t)$:
\begin{eqnarray}
&&\frac{\partial M(t)}{\partial \vec v_{\phi}(t)} = \left[\frac{\partial \vec G_{\phi}(t)}{\partial \vec v_{\phi}(t)}.\vec v_{\phi}(t) + \vec G_{\phi}(t)\right] -  \nonumber \\
&&\left[ \frac{\partial \vec R_{s}(t)}{\partial \vec v_{\phi}(t)}.\vec v_{\phi}(t) - \vec R_{s}(t)\right]=0 \label{(4)}
\end{eqnarray}
Then the velocity permitted by the instantaneous deformation $\vec v_{\phi }$ is given by:
\begin{equation}
\vec v_{\phi}(t) = -\frac{\left[ \vec G_{\phi}(t) - \vec R_{s}(t)\right]}
{\left[ \frac{\partial \vec G_{\phi}(t)}{\partial \vec v_{\phi }(t)} - \frac{\partial \vec R_{s}(t)}
{\partial \vec v_{\phi}(t)}\right]}\label{(5)}
\end{equation}
ii) for $\vec v_{s}(t') = \vec v_{\phi}(t+\tau)$:
Another differential equations of evolution of the dynamical system from Eq. (\ref{(1)}):
\begin{equation}
\frac{\partial \psi _{s}(t')}{\partial \vec v_{s}(t')} = \frac{\partial \left(\vec R_{s}(t').\vec v_{s}(t') \right)}{\partial \vec v_{s}(t')} = 0 \label{(6)}
\end{equation}
Then the velocity of formation of the dissipative pattern  is given by:
\begin{equation}
\vec v_{s}(t') =  - \frac{\vec R_{s}(t')}{\frac{\partial \vec R_{s}(t')}
{\partial \vec v_{s}}}\label{(7)}
\end{equation}
This differential equation shows that the form of the function $\vec R_{s}=u_{s}dN_{s}/dL_{s}\approx u_{s}D_{q}\left(\frac{l_{\min}}{L_{s}}\right)^{1-D_{q}}\frac{1}{l{\min}}$
is a kind of homogenous function that satisfy the Euler's theorem for fractal or multifractal patterns.
According to the fractal theory, if the homogeneity is global, i.e. $q=0$ one has a fractal, or if the homogeneity is local, i.e. $q\neq 0$, change with regions, one has a multifractal \cite{10}.

Upon receipt of this differential equations system we can get the functional dependencies of the velocities for the formation of the dissipation structure, developing the differential Eq. (\ref{(4)}))  in the following way:
\begin{equation}
\frac{\partial \vec G_{\phi }(t)}
{\partial \vec v_{\phi}(t)}\vec v_{\phi}(t) + \vec G_{\phi}(t) = \frac{\partial \vec R_{s}(t)}
{\partial \vec v_{\phi}(t)}\vec v_{\phi}(t) + \frac{\partial \vec R_{s}(t')}
{\partial \vec v_{s}(t')}\frac{\vec R_{s}(t)}
{\frac{\partial \vec R_{s}(t')}
{\partial \vec v_{s}(t')}} \label{(8)}
\end{equation}
using the Eq. (\ref{(7)}) in Eq. (\ref{(8)}) one has
\begin{equation}
v_{s}(t') = \frac{ -\vec G_{\phi}(t)}
{\frac{\partial \vec R_{s}(t')}
{\partial \vec v_{s}(t')}}\left\{1 - \frac{1}{\vec G_{\phi}(t)}\left( \frac{\partial \vec R_{s}(t)}
{\partial \vec v_{\phi}(t)} - \frac{\partial \vec G_{\phi}(t)}
{\partial \vec v_{\phi}(t)} \right)\vec v_{\phi}(t) \right\}
\label{(9)}
\end{equation}
In analogous way made before in the Eq. (\ref{(7)}) one can consider that the $\vec G_{\phi}(t)$ is too homogenous.
Therefore, introducing the following term:
\begin{equation}
\mu(t,t') =  \frac{\frac{ - \vec G_{\phi}(t)}{\vec v_{\phi}(t)}}
{\frac{\partial R_{s}(t')}{\partial \vec v_{s}(t')}}=\frac{\frac{ - \vec G_{\phi}(t)}{\vec v_{\phi}(t)}}
{\frac{\partial \left( u_{s}\vec \rho_{s}(t') \right)}{\partial \vec v_{s}(t')}}
\label{(10)}
\end{equation}
where $\mu(t,t')$ is a control parameter, one has:
\begin{equation}
\vec v_{s}(t') = \mu \vec v_{\phi}(t) \left\{ 1 - \frac{1}{\vec G_{\phi}(t)}
\left[ \frac{\partial \vec R_{s}(t)}{\partial \vec v_{\phi}(t)} -
\frac{\partial \vec G_{\phi}(t)}{\partial \vec v_{\phi}(t)} \right]\vec v_{\phi(t)} \right\}
\label{(11)}
\end{equation}
Denoting:
\begin{equation}
\vec v_{\phi \max} =
-\frac{\vec G_{\phi}(t)}
{\left[ \frac{\partial \vec G_{\phi}(t)}{\partial \vec v_{\phi}(t)} - \frac{\partial R_{s}(t)}{\partial \vec v_{\phi}(t)} \right]} =
-\frac{\vec G_{\phi}(t)}
{\left[ \frac{\partial \vec G_{\phi}(t)}{\partial \vec v_{\phi}(t)} - \frac{\partial \left(u_{s}\rho _{s}(t)\right)}{\partial
\vec v_{\phi}} \right]}
\label{(12)}
\end{equation}
from Eq. (\ref{(5)}) for $t=t'$ one gets the classical equation for the crack growth velocity:
\begin{equation}
\vec v_{\phi}(t) = \vec v_{\phi \max}\left(1 - \frac{\vec R_{s}(t)}{\vec G_{\phi}(t)} \right) \label{(13)}
\end{equation}
and finally from Eq. (\ref{(11)}) the following relation between the velocities is then supplied:
\begin{equation}
\vec v_{s}(t') = \mu(t,t') \vec v_{\phi }(t)\left(1 - \frac{\vec v_{\phi }(t)}{\vec v_{\phi \max }} \right) \label{(14)}
\end{equation}
and from Eq. (\ref{(7)}) one gets that 
\begin{equation}
\mu(t,t')=\left(\frac{\vec G_{\phi}(t)}{\vec R_{s}(t')}\right)\left( \frac{\vec v_{s}(t')}{\vec v_{\phi}(t)}\right)
\label{(15)}
\end{equation}

Therefore from Eq. (\ref{(14)}) one has the classical equation for the elastodynamic energy released rate, $\vec G (L_{\phi},v_{\phi})$ given by:
\begin{equation}
\vec G_{\phi}(L_{\phi}(t),v_{\phi}(t))=\frac{\vec R_{s}(t')}{\left(1 - \frac{\vec v_{\phi }(t)}{\vec v_{\phi \max}} \right)} \label{(16)}
\end{equation}
This equation a direct consequence of MEDP.

\subsection{\label{subsec:3}The Iterated Functions and the Logistic Map}
\begin{figure}
\includegraphics[scale=0.7]{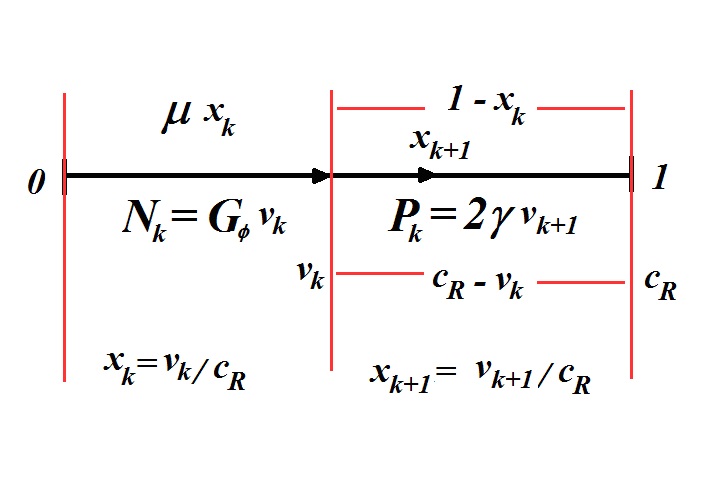}
\caption{Interpretation of the logistic equation in terms of the fast crack growth}
\label{fig:1}
\end{figure}

Considering the time delay between the velocities where $\vec v_{s}=\vec v_{\phi}(t+\tau)$, from Eq. (\ref{(14)}) one obtains the logistic equation $\vec v_{\phi}(t+\tau)=\mu \vec v_{\phi}(t)(1 - \vec v_{\phi}(t)/\vec v_{\phi \max})$. In terms of a discretized process that happens in steps of time, $\tau$, for $x_{k+1}=v_{0}(t+\tau)/c_{R}$ and $x_{k}=v_{0}(t)/c_{R}$  one has:
\begin{equation}
x_{k + 1} = \mu \left(1 - x_{k} \right)x_{k} \label{(17)}
\end{equation}
This equation shows that the energy dissipation can also be written in the form of a logistic map with the consequence that critical velocity, instability, bifurcation and crack branching, are not explained by the classical theory of fracture. The interpretation of this equation for the dynamic crack growth is shown in the Fig.~\ref{fig:1}.

In this figure, the injected flux $N_{k}=\vec G_{k}.\vec v_{k}$ interacts with the dissipated flux $P_{k}=2\gamma v_{k+1}$
in a way that the partial fraction of the velocity $\mu x_{k}$ is balanced by the remaining partial fraction of the velocity $1-x_{k}$.

\begin{figure}
\includegraphics [scale=0.45]{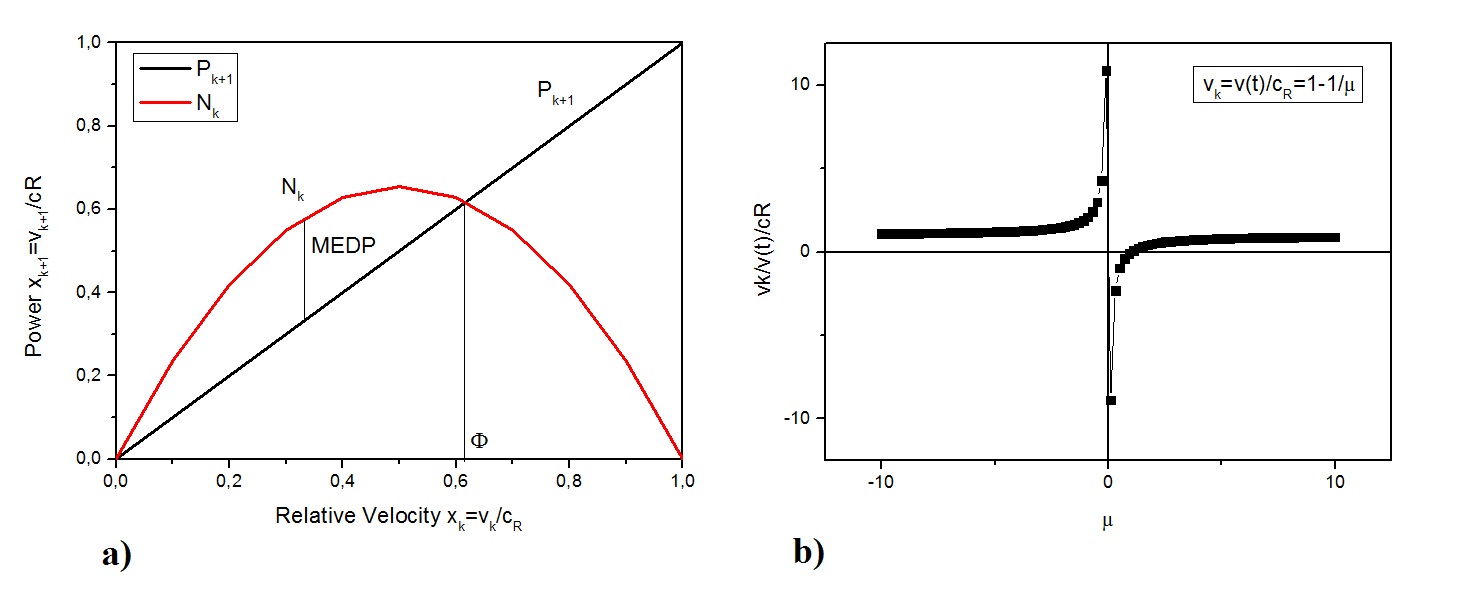}
\caption{a) Maximum Dissipation Principle in terms of iterated function of  $N_{k} = G_{\phi}.x_{k}$ versus
$P_{k+1} = R_{s}.x_{k+1}$, for logistic equation; b) Solutions spectra for logistic map in the intervals of  unstable interaction of input and output fluxes in dynamic fracture problem for $\vec v_{k} = 1 - 1/\mu$}
\label{fig:2}
\end{figure}

The energy flux to the crack tip can be represented by Fig.~\ref{fig:2}a. Observe from this that MEDP (Slepyan Principle) along with decoupling in time between the velocities provides the mathematical solution for the fracture problem as logistic iterated system of equations with a time delay. In the Fig.~\ref{fig:2}a one can also observe the point where the power and the efficiency are optimized when the Eq. (\ref{(17)}) is satisfied. The plot of iterated functions from Eq. (\ref{(17)}) is shown in the Fig.~\ref{fig:2}, and for the first iteration the solution of the logistic Eq. (\ref{(17)})) is given by the plot in the Fig.~\ref{fig:2}b.

A non-linear instability analysis can be done considering the critical points, where $x_{k+1}=x_{k}=x^{*}$ and $x^{*}=v/c_{R}$. The first critical point, is given by $dF(x^{*})/dx = 2 - \mu  < 1 \Rightarrow \mu >1$ if $1 \leqslant \mu  \leqslant 4$. Interpreting this in terms of a dynamic process of crack growth, then,
\begin{equation}
x^{*} = 1 - \frac{1}{\mu } \label{(18)}
\end{equation}
if $\mu < 1 \Rightarrow {x^*} < {\text{0 (reversible process)}}$ or if $\mu= 1 \Rightarrow {x^*} = {\text{0 (stable process)}}$ or if $\mu > 1 \Rightarrow {x^*} > {\text{0 (unstable irreversible process)}}$.

\begin{figure}
\includegraphics[scale=0.6]{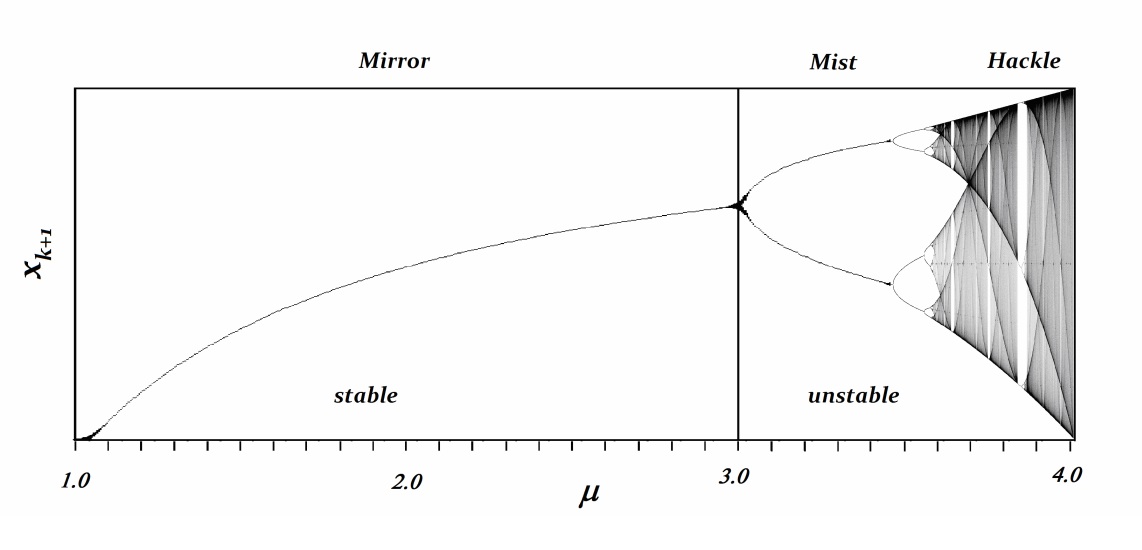}
\caption{Bifurcation map of $x_{k + 1}$ versus $\mu$ , for logistic equation}
\label{fig:3}
\end{figure}

For more feed-backed irreversible iterations a doubling period can be obtained from the recursive Eq. (\ref{(17)}). Successive iterations of this function can generate a bifurcation map as shown in the Fig.~\ref{fig:3}. This map reveals the transition from stable to unstable process for $\mu = 3$, until the system reach the chaos for a value of $3 < \mu < 4$.

The plot of the Fig.~\ref{fig:3} can represent the different aspects, such as, mirror, mist and hackle, associated to the stages of the fracture surface of brittle materials, found in testing with high velocities or high rates of loading, as PMMA, glass, etc.

\section{\label{sec:3}Comparison with Experimental Results}
Substituting all the experimental values obtained for Fineberg-Gross in the Eq. (\ref{(2)}), for the crack growth velocity of PMMA, it was found the same time delay of  $\tau=3.0\mu s$ , evidencing that this time is a characteristic for the motion in this material. The critical velocities at which the instability begins for PMMA and soda-lime glass are $340m/s$ and $1100m/s$, respectively, and the respective values of superficial Rayleigh waves velocity, $c_{R}$, in these materials are $975m/s$  and $3370m/s$. For both materials the critical values correspond approximately to $v_{c}=0.34c_{R}$. The predicted chaotic model  for the PMMA, provides a critical point at  $\mu = 3$, where the critical velocity is $v_{c}=c_{R}/3$. Comparing with the experimental results obtained by Fineberg-Gross one has an agreement about $97\%$.

\begin{figure}
\includegraphics[scale=0.45]{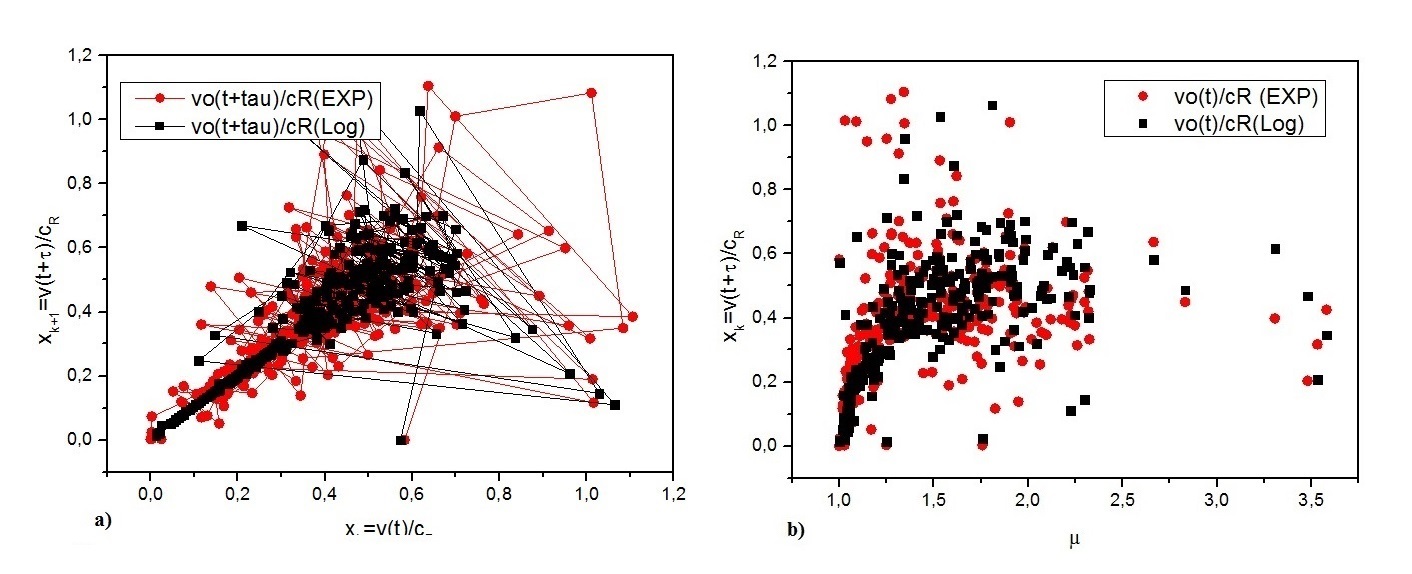}
\caption{Comparison between theoretical and experimental results; a)  The relatives velocities $x_{k + 1} = v_{\phi}(t + \tau)/c_{R}$  versus $x_{k} = v_{\phi}(t)/c_{R}$;  b)  the relative crack velocity $v_{\phi}(t+\tau)/c_{R}$ versus $\mu(t,t')$.
\label{fig:4}}
\end{figure}

The maps associated to the dissipation patterns of cracks in fast crack growth phenomena are shown in the Fig.~\ref{fig:4}a and Fig.~\ref{fig:4}b. The equation Eq. (\ref{(3)}) relates the injected power $N_{0k}=G_{\phi}x_{k}$ with the dissipated power $P_{0k+1}=2\gamma_{ef}x_{k+1}$. It was used to construct the iterated map shown in the Fig.~\ref{fig:4}. This figure was constructed in a similar manner of the illustrative Fig.~\ref{fig:2}a. From the Eq. (\ref{(3)}), the theoretical values of $x_{k+1}=v_{\phi}(t+\tau)/c_{R}$ were plotted as a function of $x_{k}=v_{0}(t)/c_{R}$, in comparison with the experimental results. Using the same values of $t$ and $\mu$ in both cases, one gets the iterated map shown in the Fig.~\ref{fig:4}a. This graph shows that theoretical and experimental values are in good agreement.

The logistic Eq. (\ref{(17)}) was used to construct the  logistic bifurcation map.
From this equation the theoretical values of $v_{\phi}(t+\tau)$ were plotted as function of $\mu(t)$,
in comparison with the  experimental results.
Using the same multiples values of $t' = t+\tau$ and $\mu$ in both cases,
the logistic bifurcation map was obtained where this graph is shown in the  Fig.~\ref{fig:4}b.
This graph shows how approximate are the theoretical with experimental values, at each iteration $k$.

\begin{figure}
\includegraphics[scale=0.45]{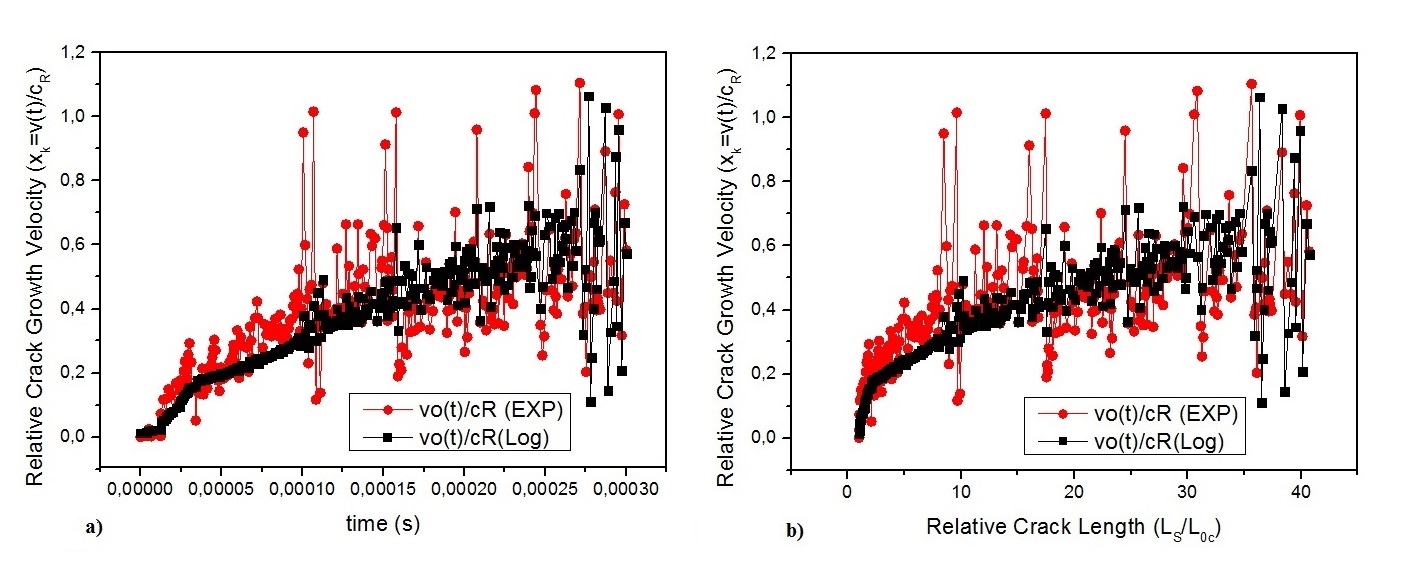}
\caption{Comparison between theoretical and experimental results; adapted from Gross, Steven Paul \cite{11}, Doctor Philosophy Dissertation, 1995. a) the relative crack velocity $v_{\phi}(t + \tau)/c_{R}$  versus time $(t)$; b) the relative crack velocity $v_{0}(t)/c_{R}$ versus relative crack length  $L_{\phi}(t)/L_{0c}$.
\label{fig:5}}
\end{figure}

The Eq. (\ref{(17)}) was used too to construct the graph shown in the  Fig.~\ref{fig:5}a.
From this equation the theoretical values of relative velocities $\frac{v_{\phi}(t+\tau)}{c_{R}}$
were plotted as function of  time $t$,
in comparison with the  experimental results. This was done using the same values of $\mu$(t) in  both cases.

The Eq. (\ref{(17)}) was used too to construct the shown in the  Fig.~\ref{fig:5}b.
From this equation those theoretical values of relatives velocities $\frac{v_{\phi}(t+\tau)}{c_{R}}$
were plotted as function of  relative crack length $L_{\phi}/L_{0c}$
in comparison with the  experimental results.
Using the same values of $t$ and $\mu$ in  both cases.
These plots show the ultimate stage of the instability reached in the Fineberg-Gross experiments.
All theses plots show the excellent agreement between theoretical and experimental results.

\section{\label{sec:4}Discussions and Conclusions}
One proposal attempts to explain the instability process on fast crack growth were made before \cite{8,9} but without any clear experimental comparison. Now in this paper, a fundamental principle was introduced, which allow us to obtain the iterated equations and enabling to foresee the dissipation pattern in a complete mathematical scenario, from non-linear dynamical equations up to the fractal bifurcation map.

The analysis of the bifurcation diagram given by the logistic map equation, shows that the way found by the material to dissipate the elastodynamic energy excess imposed on it during the fracture process is by generating alternative ways of energy dissipation through the mechanism of rough instabilities and bifurcations. Therefore, it was shown from the results obtained with this proposed model that bifurcation process is in agreement with the assertion made by Gross \cite{11}:

"The bifurcation is a response to the system being driven further from equilibrium and trying to find new ways to dissipate the forcing. This means that studying the details of the pattern can lead both to an increased understanding of the bifurcation and to the system’s dissipation process".

Since there are several  physical systems where the bifurcation happens, this can reflect a general feature of branched phenomena in nature, where  geometric patterns are generated to optimize the energy dissipation. Note that the formulated MEDP is completely general. However, this result Eq. (\ref{(17)}) follows naturally from
Eq. (\ref{(1)}) and was deduced as a special case, by taking the fracture of a semi-infinite Griffith plate \cite{8}. Therefore, it depends on the phenomenology, i.e., in which case the dynamic fracture was applied. Moreover, other iterated functions can be obtained from MEDP, as the phenomenon is generating other maps different from logistic map obtained for the case of semi-infinite Griffith plate.

With the assumption of different forms of energy input $\dot\phi \left(L_{\phi}(t),\vec v_{\phi}(t)\right)$  and other hypotheses of the dynamics mechanism provided by the system for the formation of dissipation branched patterns, given by $\psi_{s}\left(L_{s}(t'),\vec v_{s}(t')\right)$ , the  Eq. (\ref{(1)}) can give rise to different bifurcation maps; this will make the whole framework for the proposed principle much more complex than the one presented in this work. Therefore, the purpose of this work is to highlight an entire phenomenological framework that is still "untied" and illustrating it with a known logistic map that can be very well used in the analysis of dynamic fracture.

Daniel M. Dubois in 2006 \cite{12} generated and classified families of iterated functions that created maps similar to the logistic map. The study of Dubois can be used to obtain results in balancing the different dynamic systems that exhibit instability and bifurcation.

Still notice that time delay between the input and output fluxes in the system, originated the iterated equations and that on its turn gave rise to the instability by means of bifurcation maps. Since Feigenbaum constant \cite{13} is universal and common to all of bifurcation maps, one can conclude that a universality of these maps is also associated with the universality of the principle of maximum energy dissipation. As a general principle in Nature dissipation of energy  should be maximum while the excess, is minimal. Therefore, it is possible that other dissipation energy patterns can be predicted by this modified MEDP, using the same mathematical  scenario.

\section{Acknowledgments}
Lucas M\'{a}ximo Alves, thanks the 
Profs. Drs. Leonid Slepyan and Prof. Benjamin de Melo Carvalho for helpful discussions
Rui F. R. M. Lobo thanks the FCT-MCTES -Portugal through PEST (UID/EEA/00066/2013)  is gratefully acknowledge


\end{document}